\documentstyle[times,aas2pp4]{article}

\newcommand{\dexp}[1]{$\times 10^{#1}$}

\newcommand{\flx}{erg (cm$^{2}$\AA~s) $^{-1}$ {}}

\newcommand{\ea}{{\em et~al.\ }{}}
\newcommand{\msol}{M$_\odot$}

\newcommand{\eg}{{\em e.g.~}{}}

\begin{document}

\title{Ultraviolet Imaging of the z=0.23 Cluster Abell 2246} 

\author{  Robert H. Cornett\altaffilmark{1},
           Ben Dorman\altaffilmark{2,3},  
           Eric P. Smith\altaffilmark{3,7},  \\
           Michael A. Fanelli\altaffilmark{1},
           William R. Oegerle \altaffilmark{6,7},
           Ralph C. Bohlin\altaffilmark{4},  \\
          Susan G. Neff\altaffilmark{3}, 
          Robert W. O'Connell\altaffilmark{2}, 
          Morton S. Roberts\altaffilmark{5}, \\
          Andrew M. Smith\altaffilmark{3}, 
       and Theodore P. Stecher\altaffilmark{3}  }        
          
\altaffiltext{1}{Raytheon STX Corporation, Code 681, Goddard Space Flight Center,
  Greenbelt MD 20771} 
\altaffiltext{2}{University of Virginia, Astronomy Department,
  P.O. Box 3818, Charlottesville, VA 22903}
\altaffiltext{3}{Laboratory for Astronomy and Solar Physics,
  Code 680, Goddard Space Flight Center,  Greenbelt MD 20771}
\altaffiltext{4}{Space Telescope Science Institute, Homewood Campus, 
Baltimore MD 21218}
\altaffiltext{5}{NRAO, 520 Edgemont Rd., Charlottesville, VA 22903-4575}
\altaffiltext{6}{Department of Physics and Astronomy, Johns Hopkins University,
Baltimore, MD 21218} 
\altaffiltext{7}{Visiting Astronomer, Kitt Peak National Observatory (KPNO), 
which is operated by Association of Universities for Research in Astronomy
 (AURA), Inc. under contract to the National Science Foundation}


\begin{abstract}

We present deep ultraviolet observations of a field containing the cluster 
Abell 2246 (z=0.225) which provide far-ultraviolet (FUV) images of some of 
the faintest galaxies yet observed in that bandpass.  Abell 2246 lies within 
the field of view of Ultraviolet Imaging Telescope (UIT) observations of the 
quasar HS1700+64, which accumulated over 7100 seconds of UIT FUV exposure time 
during the Astro-2 mission in March 1995.  For objects found on both the FUV 
and ground-based V-band 
images, we obtain FUV ($\lambda \sim$ 1520\AA) photometry and V-band
photometry, as well as mid-UV ($\lambda \sim$ 2490\AA) photometry from 
UIT Astro-1 observations and ground-based I-band photometry. 
We find five objects in the images which are probably galaxies
at the distance of Abell 2246, with FUV magnitudes (m(FUV))
between 18.6 and 19.6, and V magnitudes between 18.4 and 19.6. We find that
their absolute FUV fluxes and colors imply strongly that they are 
luminous galaxies with significant current star formation, as well as 
some relatively recent, but not current, ($>$400 Myr ago) star formation. We 
interpret the colors of these five objects by comparing them with local objects, 
redshift-corrected template spectra and stellar population 
models, finding that they are plausibly matched by 10-Gyr-old population 
models with decaying star formation, with decay time constants in  
the range 3 Gyr $<\tau<$ 5 Gyr, with an additional color component from a 
single burst of moderate ($\sim$400-500 Myr) age. 
From derived FUV luminosities we compute current star formation 
rates.  We compare the UV properties of Abell 2246 with those of the Coma
cluster, finding that Abell 2246 has significantly more recent star formation,
consistent with the Butcher-Oemler phenomenon.


\end{abstract}

\section{Introduction}

UV images of local galaxies provide unique direct 
evidence about the morphology and quantitative nature of recent 
star formation, while UV images of objects at moderate redshifts 
(z up to $\sim$0.3) extend the limits of direct knowledge of massive 
star formation back to more remote times.  The darkness of the 
UV sky (\cite{oconnell}) further enhances observational prospects in this 
waveband. However, deep UV images of such objects are rare because 
of the dearth of instruments
with sufficient resolution and sensitivity.  In this paper we discuss deep 
observations including photometry in the far UV ($\sim$1520\AA; FUV) and 
mid UV ($\sim$2490\AA; MUV) by the Ultraviolet Imaging Telescope (UIT) of 
a field which contains the cluster Abell 2246 (\cite{abell}).  Abell 2246
is only lightly 
obscured by Galactic foreground reddening (E(B--V)=0.03) 
but has nevertheless attracted few published observations.  Cluster center,
angular size, richness, and luminosity function data for Abell 
2246 used here are from \cite{hoessel1}, \cite{hoessel2} and \cite{struble}. With a distance 
of 900 Mpc and a look-back time of 1.2 Gyr (based on z=0.225, 
for q$_{0}$ = 0.1 and H$_{0}$ = 75 km~s$^{-1}$~Mpc$^{-1}$, adopted for this 
paper) galaxies in Abell 2246 are among the most distant and faintest yet 
observed in the far ultraviolet. They provide an important stepping stone to 
understanding observations of more distant objects in longer-wavelength bands.

\section{Observations and Data Reduction}

Figure \ref{fullfield} shows the UIT FUV image, the UIT MUV image, and the 
V-band image, 
co-aligned and with the same scale, produced as described in the rest of 
this section.  North and East, and a 1\arcmin\ bar,
are shown.  Sources which are identified in this work as galaxies at the 
distance of Abell 2246 are outlined by circles and numbered as in Table~
\ref{datatab}. Other nonstellar sources found on both images
including apparent foreground galaxies and the quasar HS1700+64, are outlined
by squares.   

\begin{figure*} [pht]
\vtop{
\vglue 7.0 truein
}
\vskip 0.5truein
\caption{\protect\label{fullfield}(a) The UIT FUV image. the field of view is 20\arcmin\ across. 
North, East, and a 1\arcmin\ bar, are noted. Sources identified here as 
galaxies at the distance of Abell 2246 are outlined by circles; other 
nonstellar sources found on both images including apparent foreground 
galaxies and the quasar HS1700+46 are outlined by squares.  Source numbers 
refer to Table \protect\ref{datatab}.}   
\end{figure*}

\begin{figure*}[pht]
\figurenum{\ref{fullfield}}
\vtop{
\vglue 7.0truein
}
\vskip 0.5truein
\caption{ (b) The UIT\  MUV image, 
co-aligned and with the same scale as the FUV image; the field of view is 20\arcmin\ across. 
North, East, and a 1\arcmin\ bar, are noted. Sources identified here as 
galaxies at the distance of Abell 2246 are outlined by circles; other 
nonstellar sources found on both images including apparent foreground 
galaxies and the quasar HS1700+46 are outlined by squares.  Source numbers 
refer to Table \protect\ref{datatab}.}
\end{figure*}

\subsection{Ultraviolet Images}

The UIT images are approximately centered on the quasar HS1700+64, which was 
a high priority target for the Hopkins Ultraviolet 
Telescope (HUT) science program.  It was observed during two orbits during 
December 1990 by the Astro-1 ultraviolet instruments, and seven orbits  
during March 1995 by Astro-2.  The quasar, at z=2.743, was utilized as 
the background UV source for the measurement of \cite{davidsen} of the HeII opacity in
the intergalactic medium.  UIT and the Wisconsin Ultraviolet Photometer-
Polarimeter Experiment (WUPPE; \cite{wuppe}) were co-pointed with HUT
for these observations.  

UIT's intensified film cameras obtained a total of  
7 MUV exposures and 23 FUV exposures of a 40\arcmin\ diameter field 
surrounding the quasar, with exposure times ranging from 6 to 1947 seconds 
during the Astro missions.
Astro-1 exposures of this field were made early in the mission.  They have
have stellar full width at half-maximum (FWHM) $\sim$4\arcsec, reflecting
the image quality obtained before 
nominal pointing performance was achieved. However, the Astro-2 images 
are more typical of UIT data, having stellar PSFs with FWHM $\sim$3\arcsec.  
During Astro 2, nine images were made through UIT's broadest-band 
FUV filter, known as ``B1'', which has effective wavelength for flat spectra 
1521\AA, bandwidth 354\AA, and is blocked against Ly$\alpha$ skyglow. 
Five of these images with individual exposure times of more than 
$\sim$1000 seconds are used in this study.  No MUV images are available 
from the Astro-2 mission.  Individual 
exposures used here are listed in Table~\ref{uvobstab}.  UIT film images are
digitized, linearized, flat fielded, and absolutely calibrated.  The 
calibration and reduction procedures, combined with UIT's image intensifiers,
which eliminate reciprocity failure, produce linear, high-precision 
photometric data with typical calibration uncertainties of order 15\% for 
well-exposed sources. Details of UIT 
hardware, calibration, operations, and data reduction are in \cite{stechera} 
and \cite{stecherb}. 

\placetable{uvobstab}


\begin{deluxetable}{llrllll}
\tablecolumns{7}
\tablecaption{UIT Images \label{uvobstab}}

\tablehead{
\colhead{Image\#}  &  \colhead{Field Center} & \colhead{Exp. Time} & 
\multicolumn{2}{c}{Obs. Epoch} & 
\colhead{Filter }  &  \colhead{Flight} \\
\colhead{} & \colhead{RA (2000.0) DEC} & \colhead{(sec)} & 
\multicolumn{2}{c}{(GMT)} & \colhead{} & \colhead{} \\ }

\startdata

  FUV2015 & 17 00 45.1  +64 13 09. &  974. & 3/05/95 & 00:15 & B1
  & Astro-2   \nl
  FUV2074 & 17 00 45.1  +64 13 09. & 1210. & 3/06/95 & 00:34 & B1
  & Astro-2   \nl
  FUV2257 & 17 00 45.1  +64 13 09. & 1460. & 3/08/95 & 01:21 & B1
  & Astro-2   \nl
  FUV2260 & 17 00 45.1  +64 13 09. & 1560. & 3/08/95 & 02:50 & B1
  & Astro-2   \nl
  FUV2366 & 17 00 45.1  +64 13 09. & 1946. & 3/10/95 & 00:30 & B1
  & Astro-2   \nl
  NUV0176 & 17 00 36.7  +64 13 42. &  745. & 12/06/90 & 05:26 & A1 
  & Astro-1   \nl

\enddata

\end{deluxetable}

The FUV images listed in Table~\ref{uvobstab} were  converted to intensity 
units using the FLIGHT22 calibration constants and processing 
(\cite{stecherb}).  The images were resampled in software 
to the astrometry of the single exposure UIT FUV2366, and added to 
create a ``stacked'' image with a total exposure time of 7152 seconds. 
Co-alignment of the individual exposures was based on 
the standard astrometric solutions derived for UIT images using HST 
guide stars as fiducials (\cite{lasker}).  Typical positional errors 
resulting from this procedure, including image distortion residuals after 
correction (\cite{greason}), are 3-4\arcsec,  confirmed for these
images by the measured FWHM of $\sim$3.5\arcsec\ for stars on the stacked image.
Since the measured and expected astrometric alignment errors are only slightly 
larger than nominal single-exposure UIT values, no additional alignment 
procedures were
performed.  The central 1024$\times$1024 pixel ($\sim$20\arcmin) of the 
stacked image, which encloses both the cluster diameter of 18\arcmin\
and the field of view of our groundbased images, was extracted and 
used as the primary FUV dataset.

Quantitative experience with the photometric properties of deep co-added 
photographic 
images is limited, so we have made a detailed study of the noise and 
calibration characteristics of the stacked image.  The noise characteristics
of single UIT images as determined from measurements of flat fields 
are well defined, and are quantified by the UIT ``FFVAR'' function 
in the UIT data reduction package (\cite{stecherb}).
A ``variance image'' was created by summing the variances computed by FFVAR
for each pixel of each of the exposures used to create the stacked image. 
The stacked image's noise characteristics were also computed 
directly by measuring pixel-to-pixel variations on the image itself.
We compare the results of these noise determinations to similar quantities
derived from a single deep exposure (FUV2366) in the ``stack''. 
Table~\ref{noisetab} illustrates these results.  The square root 
of the exposure time ratio  
between the stacked image and the single image is 1.91.  The measured noise at 
sky levels (``$\sigma_{SKY}$'') obtained by 
measuring pixel-to-pixel variations on the images themselves, is 2.06 times 
larger for the single image than for the stacked image.  
In comparison the ratio of sigmas (``$\sigma_{FFVAR}$'')
computed from from the FFVAR function, \newline
$\rm \sigma_{FFVAR}(FUV2366)/\sigma_{FFVAR}(STACK) = 1.65.$
                   
The low measured sky noise, smaller than the 
exposure time ratio predicts and smaller than the calculated value, 
is probably due to smoothing induced by misalignments among the component 
exposures of the stacked image, over both fixed-pattern noise and faint 
sources.  The real improvement in 
signal-to-noise achieved by stacking is smaller than the exposure time ratio
predicts, since the photon statistics implied by scaling with exposure time 
are a best-case limit for photographic film detection and are appropriate 
only for well-exposed UIT pixels. 


\begin{deluxetable}{lll}

\tablecolumns{3}

\tablecaption{Noise Characteristics of Stacked Images \label{noisetab}}

\tablehead{
\colhead{} &  \colhead{FUV2366} & \colhead{Stacked Image} \\ } 

\startdata

  Exp. Time & 1946. & 7152. \nl

  $\sqrt{Exp.~time~ratio}$ & -- & 1.91 \nl

  $\sigma_{SKY}$ (\flx) & 3.3\dexp{-18} & 1.6\dexp{-18} \nl

  $\sigma_{FFVAR}$ (\flx) & 5.1\dexp{-18} & 3.1\dexp{-18} \nl

\enddata

\end{deluxetable}

The absolute calibration of the stacked image was verified by comparing
the flux measured in the image for the quasar HS1700+64 with that obtained 
simultaneously by HUT (\cite{davidsen}).  The stacked image's 
FUV flux for the quasar, measured in an aperture with diameter 20\arcsec\ and 
sky-subtracted, agrees within 6.5\%  
with the HUT sky-subtracted spectrum (kindly supplied by G. Kriss)  
made through a similar aperture, integrated over the UIT B1 bandpass.  This
difference is typical of comparisons of UIT's absolute calibration with
other instruments; when verified in this way, we estimate UIT's absolute
calibration uncertainty to be $\sim$10\%.   

The mid-UV (MUV) image used in this study, from a single exposure of 
745 seconds, was converted to intensity units using the FLIGHT15 calibration 
constants and processing developed for Astro 1 (\cite{stecherb}) and resampled 
in software to the astrometry of image FUV2366.  The resolution of this image 
as determined
by the stellar PSF FWHM is $\sim$4\arcsec\ compared to the usual MUV image 
value of 2.7\arcsec\ for images with nominal pointing quality.  Because the 
bandwidth of the MUV ``A1'' filter is 3.24 times that of the ``B1'' 
filter, and because of a known, quantified decrease in UIT sensitivity for long 
exposures during
the Astro-2 mission, the expected sensitivity limit of the MUV image is 
only 0.4 magnitudes brighter than that of the stacked FUV image.  However, the
MUV image is ``confusion limited'' at magnitude m(MUV)$\sim$20.0 due to 
the high density of noise spikes and the high frequency of spurious sources
against the sky background.  For this reason we do not accept the 
presence of a source on the MUV image as sufficient confirmation for 
detection, and include as sources only objects that appear on 
both the FUV and the V-band images.  

No independent check of MUV absolute calibration for this image is 
available because no simultaneous MUV observations of the quasar HS1700+64 
were made and no suitable IUE comparison sources are present in the 
UIT field of view. For such cases we estimate the uncertainty in the UIT 
absolute calibration to be $\sim$15\%.

\subsection{Optical-band Images}

The optical-band images, originally obtained in the course of a program to 
investigate star formation in Abell clusters,  were made with the KPNO 
0.9m telescope on 1994 July 9 using the the 2048 $\times$ 2048 T2KA CCD, 
which has 0.68\arcsec\ pixels. The 23\arcmin\ field was centered at 
RA=17$^{h}$ 01$^{m}$ 14.7$^{s}$, DEC=64$^{\circ}$ 11\arcmin45.1\arcsec\
(J2000.0).

\begin{figure*} [pht]
\figurenum{\ref{fullfield}}
\vglue 7.0truein
\vskip 0.5truein
\caption{(c) The CTIO V image. the field of view is 20\arcmin\ across. 
North, East, and a 1\arcmin\ bar, are noted. Sources identified here as 
galaxies at the distance of Abell 2246 are outlined by circles; other 
nonstellar sources found on both images including apparent foreground 
galaxies and the quasar HS1700+46 are outlined by squares.  Source numbers 
refer to Table \protect\ref{datatab}.}   
\end{figure*}

The standard UVBRI filter set was used to make broadband images for 
photometry of cluster members; those with blue colors (B--V $<$ 0.8) were 
selected as candidates for later WIYN/HYDRA spectroscopy.  
Poor seeing caused the stellar FWHM in the Abell 2246 exposures to be 
2.6\arcsec. 

V- and I-band images are used in this 
work.  The total exposure times for V-band and I-band images were each 2700 
sec, with three 900-sec exposures combined to produce the final images. 
Data were calibrated by comparing data from images of M92 made during the 
same night with published magnitudes.   

\subsection{Source Detection and Photometry}

The V-band image was used as a guide to locating galaxies in the field.  
First, non-stellar sources were found by eye on the logarithmically stretched
V-band image.  The locations of V-band objects on the the stacked FUV image 
were then searched by eye for sources within $\sim$4\arcsec.  Apparent FUV
sources were confirmed and located by centroiding (that is, by
computing the flux-weighted mean position). Data are in 
Table~\ref{datatab}, which includes positions, the distance (``Dist.'')
from the nominal cluster center in arcminutes, magnitudes in the UIT 
FUV ($\lambda$ 1520) and MUV ($\lambda$ 2490) filters, and V and I magnitudes.
Seven FUV sources 
(Nos. 1-7 in Table~\ref{datatab}) corresponding to V-band sources were 
located using this procedure.    

In addition, the automated extended-source detection and photometry package 
FOCAS (\cite{focas}) was run on the stacked image for comparison.  FOCAS 
detected a total of 79 sources including stellar ones, but did not find 
sources 4,6, and 7 in Table~\ref{datatab}, probably because they are
diffuse.  Subsequent inspection of the V-band image at the 
locations of FOCAS-found FUV  sources confirmed 6 additional
objects, numbered 8-13 in Table~\ref{datatab}.  
Of these only source 8 appears non-stellar on the V-band image.  
All 13 sources on the stacked FUV image 
were also found by eye on the MUV image. Observed positions, magnitudes,
and V-band appearance are given in the Table.



\begin{deluxetable}{llrrrrrr}

\tablecolumns{8}
\tablecaption{Photometric Results \tablenotemark{a}\label{datatab}}

\tablehead{
\colhead{Source} &
\colhead{Position} & \colhead{Dist.\tablenotemark{b}} & 
\colhead{m(FUV)\tablenotemark{c}} &  
\colhead{m(MUV)\tablenotemark{d}} & 
 \colhead{V\tablenotemark{e}} & \colhead{I\tablenotemark{e}} & \colhead{Notes\tablenotemark{f}}  \\
\colhead{\#} & 
\colhead{RA (2000.0) DEC} & \colhead{\arcmin} & \colhead{mag} & \colhead{mag}
& \colhead{mag} & \colhead{mag} & \colhead{}  \\ }

\startdata

 1 & 17 02 08.6  +64 12 20.1 & 9.2 & 18.27$\pm$0.14 & 18.44$\pm$0.16 & 16.66 & 16.53 & FG \nl
 2 & 17 01 00.5  +64 12  8.9 & 1.9 & 17.12$\pm$0.10 & 17.75$\pm$0.16 & 16.32 & 16.86 & QSO \nl
 3 & 17 01 06.3  +64 14  5.0 & 2.8 & 18.85$\pm$0.21 & 19.04$\pm$0.17 & 18.41 & 18.73 & E,S? \nl
 4 & 17 00 38.0  +64 13 46.9 & 1.2 & 19.35$\pm$0.33 & 19.51$\pm$0.20 & 18.38 & 18.37 & FG \nl
 5 & 17 00 31.4  +64 14 42.6 & 2.4 & 18.61$\pm$0.17 & 19.11$\pm$0.17 & 18.61 & 18.97 & E,S? \nl
 6 & 17 00 35.3  +64 15 21.1 & 2.8 & 19.62$\pm$0.43 & 19.86$\pm$0.23 & 18.97 & 18.99 & E? \nl
 7 & 16 59 31.8  +64 11 20.1 & 8.0 & 18.99$\pm$0.23 & 19.45$\pm$0.19 & 18.67 & 18.96 & E? \nl
 8 & 17 00 54.6  +64 04  3.5 & 8.7 & 19.43$\pm$0.34 & 19.90$\pm$0.25 & 19.63 & 19.79 & E,S?  \nl
 9 & 17 00 34.4  +64 04 45.7 & 8.0 & 19.72$\pm$0.48 & 19.08$\pm$0.17 & 15.98 & 16.41 & *  \nl
10 & 17 00 31.7  +64 07 58.5 & 4.9 & 19.72$\pm$0.47 & 21.67$\pm$1.03 & 21.11 & 19.34 & ? \nl
11 & 17 01 15.0  +64 17  0.6 & 5.5 & 19.34$\pm$0.31 & 20.10$\pm$0.27 & 19.00 & 18.11 & * \nl
12 & 17 00 38.7  +64 22 14.9 & 9.5 & 21.17$\pm$1.26 & 21.19$\pm$0.69 & 17.76 & 18.17 & ? \nl
13 & 17 00 24.1  +64 03 19.4 & 9.6 & 19.71$\pm$0.47 & 23.09$\pm$3.94 & 19.63 & 17.86 & * \nl
\enddata

\tablenotetext{a}{Not corrected for foreground extinction} 
\tablenotetext{b}{From nominal cluster center}
\tablenotetext{c}{1520~\AA}
\tablenotetext{d}{2490~\AA}
\tablenotetext{e}{Uncertainty, dominated by standard star measurements, is 0.05 mag}
\tablenotetext{f}{From appearance on V-band image: FG=foreground galaxy; 
QSO=quasar; E=elliptical galaxy; S=spiral galaxy; *=star}
\end{deluxetable}

Aperture photometry was performed on 
the images as follows; slightly different techniques were used for the 
UV and optical bands.  For the UV images, a center pixel for each 
source was chosen, either by centroiding the source or, when centroiding 
failed, using the peak pixel.  Aperture photometry in a 
circular aperture with radius 3.4\arcsec\ (11 kpc at z=0.225) was performed 
centered at that pixel location.  A local sky was subtracted, measured 
using the mean sky value in an annulus between radii of 22 and 33\arcsec, 
after removal of outliers. (A mean sky value is used for UIT images because
the nonlinear density-to-intensity conversion produces a skewed pixel value 
histogram
which does not lend itself to mode or median statistical techniques.)  The
variance in the UIT fluxes was computed conventionally, treating the
source pixels as individual measurements of different quantities, and the 
sky pixels as individual and independent measurements of a single quantity. 
Because photographic noise increases rapidly at small
signals, sky measurement dominates the UV flux 
uncertainties.  Therefore the random uncertainties in flux units for all 
sources in each UV band are approximately equal ($\sim$1.6\dexp{-17} \flx 
in the FUV, corresponding to a source of m(FUV)$\sim$ 20.9; and 
$\sim$7\dexp{-18} \flx in the MUV corresponding to m(MUV)$\sim$ 21.8).  
Uncertainties shown in Table~\ref{datatab} for the FUV and MUV magnitudes
include random error computed in this way as well as absolute calibration
uncertainties of 10\% in the FUV and 15\% in the MUV.  For the optical-band 
images standard FOCAS aperture photometry procedures, including sky 
determination and subtraction, were used, with aperture sizes matched 
to those used in the UV.  Photometric uncertainties in the optical-band
data are dominated by the reference-star uncertainties, which we estimate
at 5\%.

\section{Discussion}

Abell 2246, centered at RA=17$^{h}$ 00$^{m}$ 44$^{s}$, 
DEC=64$^{\circ}$ 12.7 \arcmin\ (J2000.0)  with a diameter of 18\arcmin, 
is completely contained within the extracted UIT sub-images centered on 
HS1700+64 as well as within the optical-band 
images. The objects 3,5,6,7, and 8 in Table~\ref{datatab} are likely 
member galaxies of Abell 2246 from several lines of argument.  
First, they appear non-stellar on the V- and 
I-band images, with sizes corresponding to linear extents of
a few tens of kpc at z=0.225. Second, their foreground- and redshift-corrected 
(FUV--V) colors, in the range -0.36 to +0.49, are similar to those of 
well-known local, bright star-forming galaxies (see Figure~\ref{cmdiag} 
and Table~\ref{abonly}, where derived FUV absolute magnitudes, UV colors, 
and star formation rates for these five galaxies are presented.).  
Third, their UV luminosities, if they are at the Abell 2246 distance, 
are approximately those of nearby UV-luminous galaxies.



\begin{deluxetable}{lrrrr}

\tablecolumns{5}

\tablecaption{Intrinsic FUV Properties of Abell 2246 Galaxies \tablenotemark{a} 
\label{abonly}}

\tablehead{
\colhead{Source} &
\colhead{M(FUV)} & \colhead{FUV--V} & \colhead{MUV--V} & \colhead{SFR} \\
\colhead{\#} & \colhead{} & \colhead{} & \colhead{} 
 & \colhead{\msol yr$^{-1}$} \\ }

\startdata

 3 & -21.20 & 0.28 & 0.49  & 2.9 \nl
 5 & -21.44 & -0.16 & 0.36 & 3.7 \nl
 6 & -20.43 & 0.49 &  0.75 & 1.4 \nl
 7 & -21.06 & 0.16 & 0.64  & 2.6 \nl
 8 & -20.62 & -0.36 & 0.13 & 1.7 \nl

\enddata

\tablenotetext{a}{After correction for Galactic foreground (E(B--V)=0.03) and 
D.M.=39.8}

\end{deluxetable}

Redshifts are necessary to confirm certainly that these five objects are not
foreground galaxies.  However two points very much strengthen the case for
this. From the UIT data compiled by \cite{fanelli}, the vast majority of 
field candidates, with ``normal'' FUV fluxes (M(FUV)$>$-18.5) 
have significantly lower FUV surface brighnesses.
Thus they cannot be placed in front of Abell 2246 so that both their 
diameters and magnitudes scale to the observed values.   
Nearly all elliptical galaxies (\eg all in Virgo) as well as 
a large fraction of spirals (\eg M81 and NGC 2903) are eliminated;  
high-luminosity objects which are allowed by this criterion must have 
DM$>$37.5, within a factor of 3 of the distance of Abell 2246. 

The strongest argument against foreground interlopers, however, is the
rarity of field galaxies, based on the very low frequency of {\em any} 
extragalactic FUV sources seen in this (1520\AA) bandpass in 
the complete UIT Astro-2 dataset. This bandpass excludes light longward
of 1800\AA\ and is therefore not sensitive to stellar populations 
characterized by stars cooler than T$_{eff}$=9000K, as discussed below.
Astro-2/UIT imaged 42 40-arcmin diameter 
extragalactic fields (that is, with $|b|>\sim$30$^{\circ}$ and excluding large 
foreground objects such as Galactic clusters and Local Group galaxies) with 
exposure times longer than 1000 seconds, with a conservative limiting 
magnitude of m(FUV)$\sim$18.5.  This image set was searched at the 
positions of all known extragalactic objects from the catalogs of 
\cite{devaucouleurs}, \cite{paturel},
\cite{huchra}, \cite{nilson}, \cite{veroncetty}, and the IRAS Point Source 
Catalog. After excluding each field's principal target and 9 known Coma 
Cluster members, a total of 16 objects were found and identified.  
Therefore, the rate of detection of serendipitous but catalogued FUV 
sources is about 0.4 objects per 40 arcmin UIT field, or 0.1 objects per 
(20-arcmin) Abell 2246 field, to m(FUV)=18.5.  We use the mid-UV 
($\sim$2000\AA) luminosity function of \cite{milliard} for background galaxies 
to scale by 1.2 magnitudes to our Abell 2246 field limit of m(FUV)=19.7, 
obtaining 0.6 possible foreground or background object per Abell 2246-sized 
field, compared with five detected objects.  Hence, the extreme rarity 
of serendipitous extragalactic FUV sources strongly supports identifying 
these objects as members of Abell 2246. 

Figure \ref{mosaic} shows $\sim$1\arcmin\ subimages of the five sources in 
Table \ref{abonly} identified here as galaxies at the distance of 
Abell 2246.  The subimages, in the V, MUV and FUV bands respectively, have 
the the same orientation as Figure \ref{fullfield}.  The objects are 
clearly not point sources, although the amount of 
morphological information available from the images is limited.

\begin{figure*}[hp]
\begin{center}
\epsfysize=6.0truein
\vglue 7.5truein
\end{center}
\caption{Subimages of the five sources 3,5,6,7 and 8 in Table \ref{datatab} 
which we identify in this work as galaxies at the distance of Abell
2246, in the FUV (left) MUV (center) and V (right) bands respectively.  
Orientation is the same as in \ref{fullfield}, and the angular scale is
shown by a 10\arcsec\ bar. 
The distance and surface brightnesses of the galaxies limit morphological
information in the images, although the objects are clearly not point sources.  
\label{mosaic}}
\end{figure*}

Assuming a distance modulus of 39.8 mag (based on z=0.225; \cite{struble}), 
we determine absolute FUV magnitudes
(corrected for Galactic foreground extinction using E(B--V)=0.03 and 
the extinction law of \cite{extinction}) in the range from --21.4 to --20.4 
(Table~\ref{abonly}). A (FUV--V), FUV color magnitude diagram for Abell 2246 
galaxies and well-known local 
objects (\cite{fanelli}) is shown in Figure~\ref{cmdiag}. In this and 
subsequent plots all observations have been corrected for Galactic 
foreground extinction.  (Note that while the color error bars 
from Table~\ref{abonly} would be a large fraction of the plot's color scale, the
plot's entire range covers only the colors of stellar spectral types B9 through 
early A, showing the leverage of the FUV bandpass in this spectral range.)      
The effect of K corrections is shown by arrows, which display the changes
in (FUV--V) color and FUV magnitude obtained by redshifting representative
galaxy and starburst spectra. To compute these vectors we
have used the template spectra of \cite{kinney}, which provide 
spectral information across the wavelength range from 1200 to 10000 \AA.  
(In this paper we use ``K correction'' to mean a transformation to rest-frame
quantities, and ``redshift correction'' to mean its inverse, a transformation 
to observed quantities).  The vectors shown are the redshift
corrections (z=0.225) of the Sc galaxy template, the mean redshift 
correction of the Sb and Sc 
galaxy templates, and the mean redshift correction of starburst templates of 
all internal reddenings, of \cite{kinney}.  (These  redshift correction 
vectors are intended to show displacements only; their locations are 
arbitrary). The observed cluster galaxies evidently have color-magnitude loci which lie within the range of 
luminous blue local objects.  However, K corrections are 
significant.  For example, if all the Abell 2246 objects are Sc galaxies, some
are intrinisically brighter in the FUV than M101, a luminosity class I Sc.

\begin{figure}
\plotone{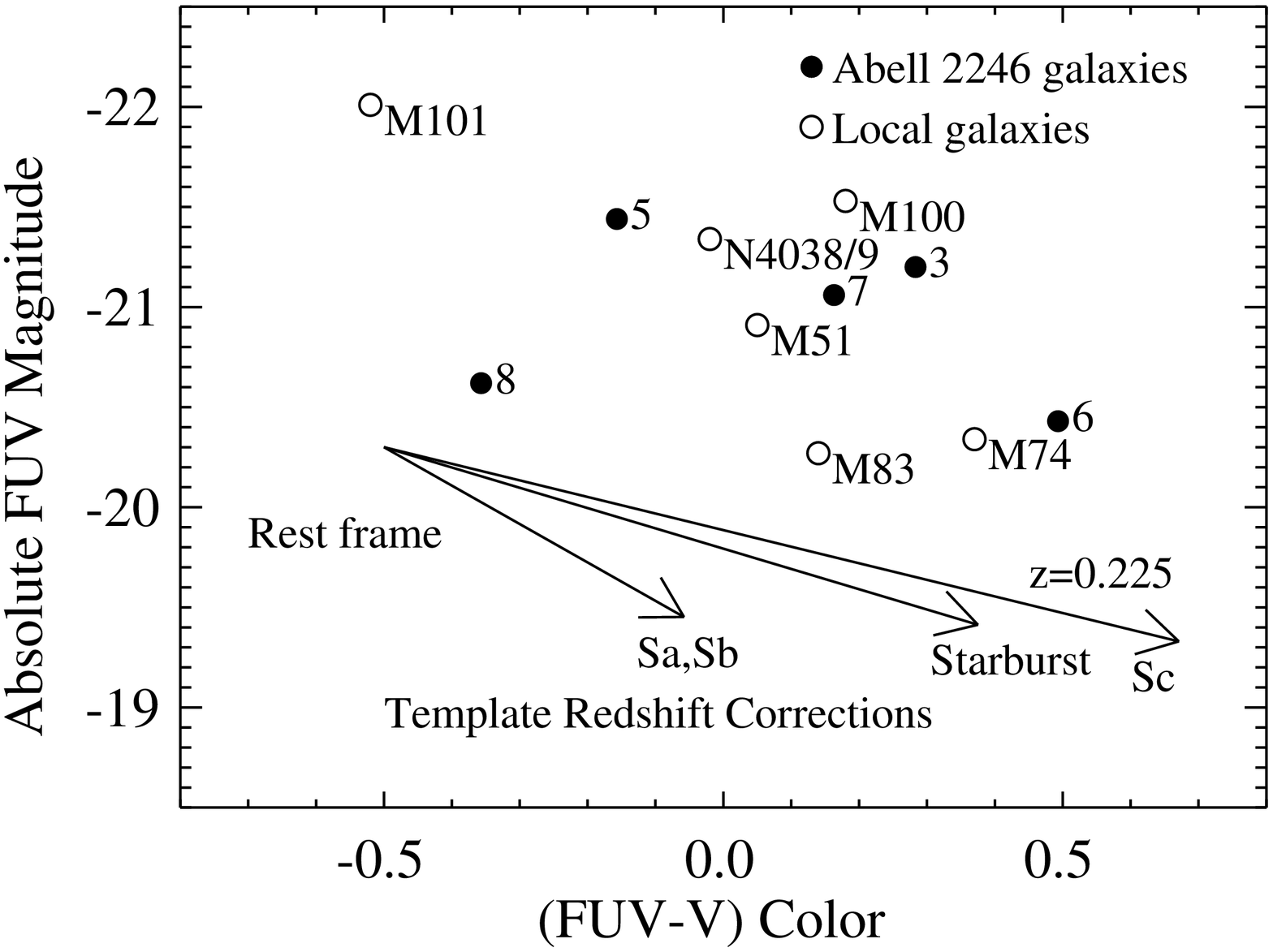}
\caption{A (FUV--V),V color absolute-magnitude diagram for Abell 2246 
(filled circles) and well-known local objects (open circles, labeled). All 
observations have been corrected for Galactic foreground extinction.  
(Note that while the color error bars 
from Table~\protect\ref{abonly} are a large fraction of the plot's color scale, the
plot's entire range covers only the colors of stellar spectral types B9 through 
early A, showing the leverage of the FUV bandpass in this spectral range.)  
Without redshift corrections,  the observed cluster galaxies 
have FUV colors and magnitudes  which lie within the range 
covered by local objects.  Redshift-correction vectors for template spectra 
of Sa and Sb galaxies (averaged), starbursts with a 
range of internal reddenings (averaged), and Sc galaxies are also shown;
appropriate  redshift corrections are evidently significant. The 
vectors are intended to show displacements only; their locations are arbitrary.
\label{cmdiag}}
\end{figure}

\begin{figure}
\plotone{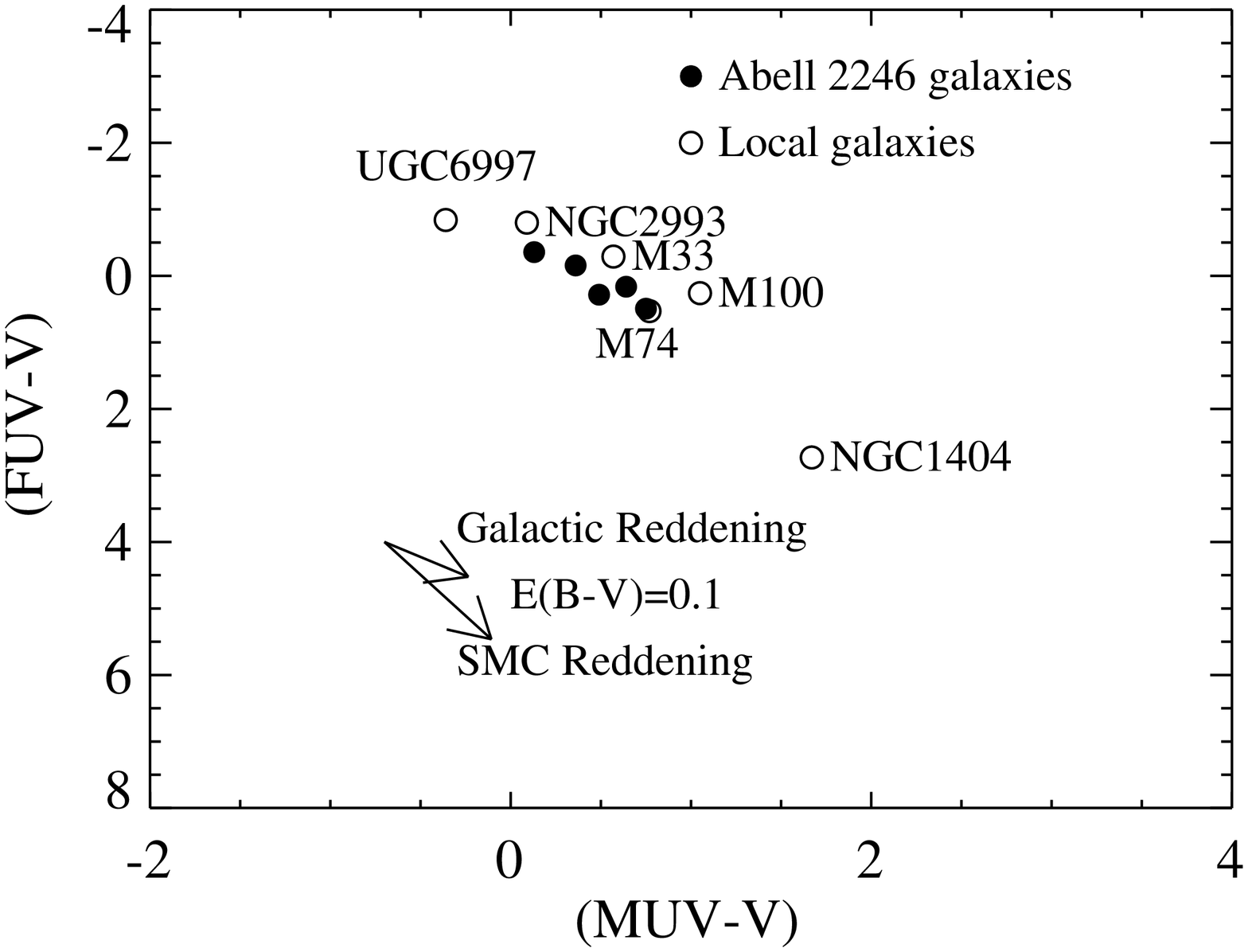}
\caption{A comparison of the (MUV--V,FUV--V) colors for Abell 2246 
galaxies (filled circles) with those derived from total magnitudes of 
observed local galaxies (open circles, labelled). All objects are corrected 
for foreground extinction; the potential effect of internal ``screen'' 
reddening for E(B--V)=0.1 for Galactic and SMC extinction laws is shown
by arrows.  Abell 2246 galaxies observed at z=0.225 have 
slightly redder FUV--V colors values than the rest-frame colors of 
the bluest star-forming systems that are plotted here.  \label{colorcoloro}}
\end{figure}

Figure~\ref{colorcoloro} compares the (MUV--V, FUV--V) colors of Abell 2246
galaxies with the integrated colors of local galaxies 
(labelled).  The Abell 2246 objects, 
observed at z=0.225, have slightly redder FUV--V colors than the rest-frame 
values of the bluest star-forming systems that are plotted here.  For 
comparison, the colors of individual annuli from M33 and M74 ({\em cf} 
Figure 7 in \cite{cornett}) form a linear distribution that runs
approximately from the locations of NGC1404 to NGC2993 on this plot, 
with colors from the outermost regions of these galaxies near those of 
NGC2993. 

\begin{figure}
\plotone{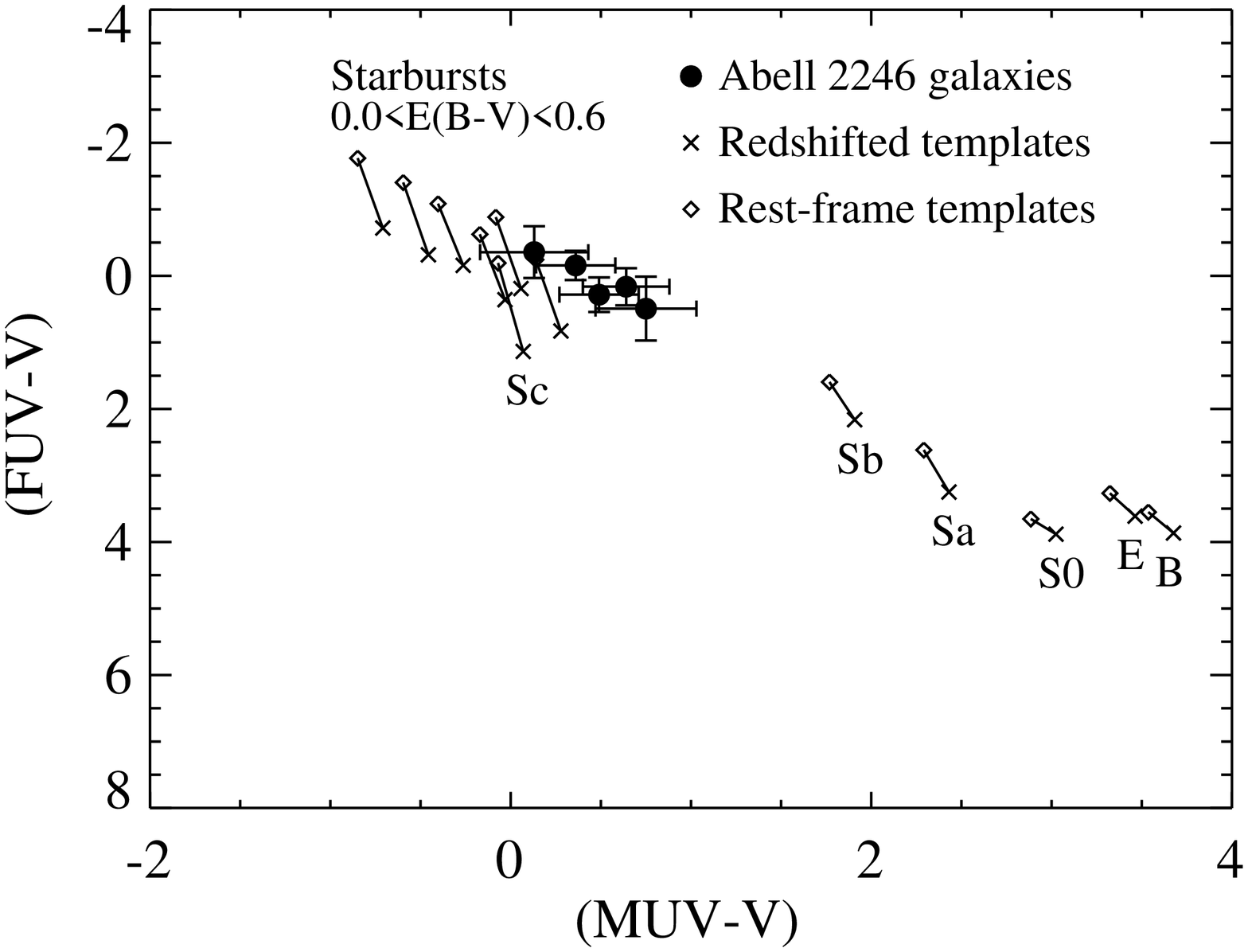} 
\caption{A comparison of the Abell 2246 data from Figure~\ref{colorcoloro}
with observation-based template galaxy spectra.  Diamonds, from upper left
to lower right,  are the rest-frame colors of starbursts with 
small (E(B--V)$<$0.1) to large (E(B--V)$\sim$0.5) internal reddening, through  
galaxies of increasingly earlier (``Sc, Sb, Sa, S0, E'') type, to those of
a bulge (``B'') population.  Template colors are also plotted as 
redshifted to z=0.225 (crosses), connected to  
corresponding rest-frame template colors.  Error bars shown here are 
appropriate for all color-color diagrams in this paper. \label{colorcolort}}
\end{figure}

\begin{figure}
\plotone{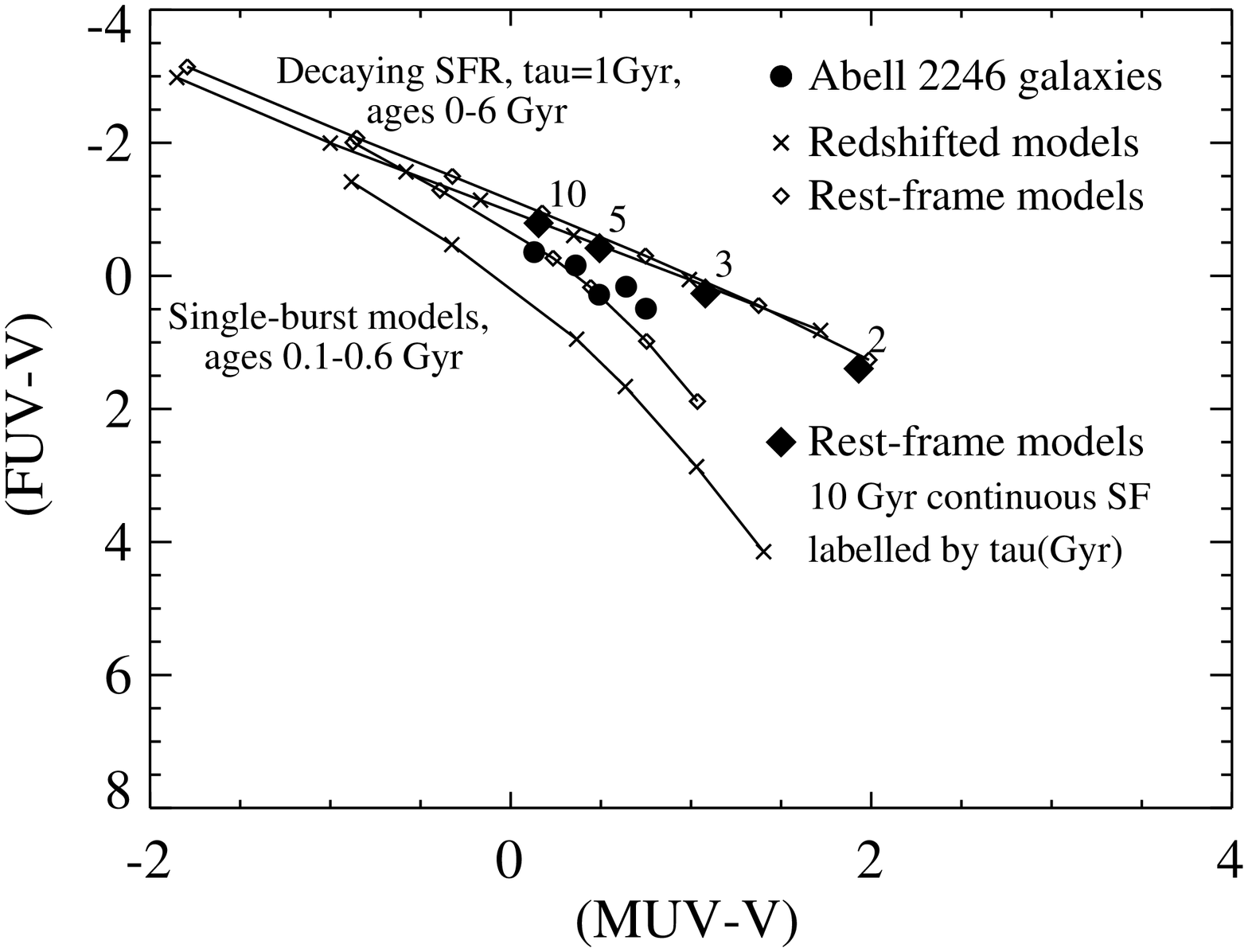}
\caption{\small{The observed MUV--V, FUV--V color-color data from 
Figure~\protect\ref{colorcoloro} displayed with the rest-frame (diamonds) and 
redshifted (crosses) evolutionary tracks of selected stellar population models.
All models have IMF slopes --1.35.  The lower pair of model curves 
represent aging single-burst star formation, with ages ranging from age .1 Gyr
(in the upper left of the diagram) to 0.6 Gyr (lower right) in $\sim$0.1 Gyr
increments.  The upper pair of curves is a continuous-star-formation model, with
a star formation rate which decreases exponentially with time constant 
$\tau$=1 Gyr, and ages from $\sim$0-6 Gyr in 1 Gyr increments.  The observed 
galaxies show strong evidence of recent ($<$400 Myr) star formation.  
Both rest-frame and redshifted model colors 
are evenly spaced with time in (FUV--V), showing a simple 
power-law increase to redder color values.  This reflects the exponential 
fall-off in numbers
of the hottest stars (which dominate the FUV bandpass) caused by the 
decaying star formation rate; the strictly straight-line relationship 
of the redshifted model is caused by the bluer bandpass of observation 
which samples only extremely hot stars. The filled diamond symbols represent 
the rest-frame colors of models of 10-Gyr-old stellar 
populations with different SFR decay times, simulating a galaxy with a 
long history of star formation. After redshift 
correction, the best model for Abell 2246 galaxy colors 
evidently has star formation decay timescales of 3 Gyr$<\tau<$5 Gyr.   
This results in models which are $\sim$0.5 mag too blue in (FUV--V); 
however, the single-burst model loci show that contributions from a single
recent burst can add MUV and V flux without significant FUV. \label{colorcolorm}}
}
\end{figure}

Figure~\ref{colorcolort} compares the (MUV--V,FUV--V) colors with data from 
the individual template galaxy and population spectra of \cite{kinney},
also shown as averages in Figure~\ref{cmdiag}.  (This Figure has
error bars appropriate for Figures~\ref{colorcoloro},\ref{colorcolort}, 
and \ref{colorcolorm}).  The template spectra are 
constructed from composites of ground-based and IUE spectra of 
individual galaxies made through apertures matched to the  
$\sim$ 10$\times$20 \arcsec\ large IUE aperture. The colors 
plotted, from upper left to lower right, range from those of composite 
starburst galaxies 
with small (E(B--V)$<$0.1) to large (E(B--V)$\sim$0.5) internal reddening as 
measured by Balmer decrements, through  
galaxies of increasingly earlier (``Sc, Sb, Sa, S0, E'') type, to those of
a bulge (``B'') population.  Each template color is plotted as viewed in 
its rest frame (diamonds) and redshifted to z=0.225 (crosses). 
The template starburst and Sc colors as well as the observed galaxy colors 
are in good agreement with bluer examples from the same bands from 
disk annuli in the local spiral galaxies M74, M81, and M33 (\cite{cornett}).

We have also computed and redshifted the spectral energy distributions 
(SEDs) of selected stellar population
models of \cite{bruzual}, shown in Figure~\ref{colorcolorm}.  
The lower pair of curves shows single-burst star-formation models 
(that is, which have no star formation after age $\sim$ 0), having solar 
metallicity, an initial mass function (IMF) slope of --1.35, and a 
range of ages.  The evolutionary tracks of those
models as viewed in their rest frame (diamonds) and with a redshift of
z=0.225 (crosses) are shown, along with the 
colors of the observed cluster galaxies
(filled circles).  The model starbursts shown range from age 0.1 Gyr 
(upper left of the diagram) to 0.6 Gyr (lower right) in $\sim$0.1 Gyr 
increments.  Evidently the observed galaxies are $\sim$1 mag bluer in 
(FUV--V) than redshifted single-burst models of the observed (MUV--V). 

The upper pair of curves in Figure \ref{colorcolorm} shows the effect of 
longer-term evolution.  Those curves display colors from a decaying 
continuous star-formation model of \cite{bruzual}, with a star formation decay 
rate time constant of
$\tau$=1 Gyr, with age increasing from $\sim$0 to 6 Gyr from upper left to
lower right.  This model, which contains both younger and older 
stars than single-burst models, produces (FUV--V) colors which are 
up to $\sim$0.7 mag bluer than observed galaxies in Abell 2246, for the 
observed (MUV--V).  

The models' behavior in the color-color plane of Figure~\ref{colorcolorm}
may be explained as follows. 
For any reasonable IMF, the UIT FUV bandpass is dominated by the hottest 
main sequence stars.  Redshifting the spectrum to z=0.225 accentuates this, 
shifting the effective bandpass of observation to shorter $\lambda$ 
(1060-1460\AA).  Therefore
the evolution of FUV flux, and (FUV--V) color, tracks changes in the total 
flux of the hottest main sequence stars.  This is borne out in the 
``Decaying SFR'' models: both the rest-frame and redshifted colors 
are evenly spaced with time in (FUV--V), showing a simple 
power-law decrease.  This behavior reflects the exponential fall-off in numbers
of the hottest stars with decaying star formation rate.  The color decay is 
nearly linear in this log-log plot because radiation at shorter (K-corrected) 
wavelengths selects the hottest stars even more stringently than the 
uncorrected bandpass. Single-burst models become redder in (FUV--V) more rapidly than ``Decaying 
SFR'' models, because the upper end of the decaying main sequence becomes 
cooler and redder, not just less populous.  The bend in the evolutionary 
curve of the single-burst models, which strongly resembles the locus of the 
colors of stellar atmospheres of a wide range of temperatures in this plane, 
is caused by the rapid fall-off of sensitivity of the FUV bandpass to cooler 
stars: at about $T_{\rm eff}=9000{\rm K}$, the peak of those stars' UV continuum 
moves longward of the FUV long-wavelength cutoff at $\sim$1800\AA.  Stars 
much cooler than this contribute significantly to MUV flux, but not to FUV.  
The location of the observed objects at significantly bluer 
(FUV--V) than redshifted single-burst models implies the existence of star 
formation more recent than the burst's age ($\sim$400-500 Myr) in the observed 
galaxies.  

The effects of redshift in Figure~\ref{colorcolorm} are of special interest.
For SEDs with current star formation and therefore a significant hot-star
component, correction to z=0.225 moves the models to bluer colors in both
(FUV--V) and (MUV--V), {\em upward} along the aging line.  In such SEDs both the 
FUV and MUV flux increase relative to V with redshift as more of the 
rising part of 
the hot-star continuum is shifted into those bandpasses.  This is opposite 
to the cases of the templates (Figure~\ref{colorcolort}) and single-burst 
models, whose SEDs lack any contributions from the hottest stars.

The filled diamond symbols represent the rest-frame colors of models 
of 10-Gyr-old stellar populations with different SFR decay times 
(\cite{cornett}).  They simulate a galaxy with a long history of star 
formation as well as strong current star formation, and may be used to 
constrain the time dependence of that star formation.  Since these model
SEDs contain important contribution from the hottest stars, redshift
corrections will move their colors upward and to the left along their 
locus, shortening their apparent SF decay timescale.  After this redshift 
correction, models consistent with Abell 2246 galaxy colors have star 
formation decay timescales of 3 Gyr$<\tau<$5 Gyr.  The models are 
still $\sim$0.5 mag too blue in (FUV--V); however, adding single-burst models 
with late A stars as their brightest component will add V and MUV flux, 
but not FUV. 

From the above discussion it is evident that a combination of 
past star formation, as represented by single-burst models,
and current star formation, as included in the models of \cite{bruzual},
can produce the colors observed in Abell 2246.  The 10-Gyr-old decaying-SFR
population with 3 Gyr$<\tau<$5 Gyr overestimates the proportion of current 
star formation, with resulting (FUV--V) colors that are too blue.  Conversely, a single-burst 
population of age $\sim$400-500 Myr is too red in (FUV--V), and implies 
that more recent star formation is required.  We conclude that the observed 
Abell 2246 galaxies a) are UV-luminous, with some as bright in the FUV as 
M101, depending on the assumed spectrum and resulting redshift correction.
b) have FUV--V colors that show clear evidence of the 
flux of relatively young ($<$100 Myr) stellar populations, and are 
undergoing significant current star formation; but 
c) like observed late-type spiral galaxies, probably require flux 
contributed by stars of a range (up to $>$ 400 Myr) of ages.     

This mixture of populations
is like  that of the objects known as ``E+A'' or ``poststarburst'' galaxies
identified by \cite{dressler}.  Spectra of these objects are thought 
to show evidence of recent (10$^{8}$-10$^{9}$ yr preceding the rest-frame
epoch), vigorous star formation which has since decreased dramatically.  
\cite{liu} model optical-band spectrophotometry of eight such objects
having 0.088$<$z$<$0.545.  Their models are constructed from a linear
combination of two spectral components: a composite spectrum of either an
E/S0 or Sbc galaxy template; and a starburst with an age in the range
10$^{8}$ to 2 $\times10^{9}$ yr.  Figures~\ref{colorcolort} and 
\ref{colorcolorm} show that it is likely that such a combination
would produce UV colors like those of Abell 2246 galaxies as well.  The red
colors of all templates except the ``Sc'' example in    
Figure~\ref{colorcolort}, compared to those of Abell 2246 galaxies, make 
it clear that a late-type underlying galaxy is required to model these 
FUV-bright objects.  

The current FUV flux, a direct tracer of 
massive stars and therefore of recent star formation, permits estimating 
rates for formation of massive (M $>$ 5 \msol) stars.  Following 
\cite{lequeux}, we use stellar evolutionary models of \cite{schaller}, 
\cite{schaerer}, and \cite{charbonnel} with the stellar atmosphere models
of \cite{kurucz}.  For comparison with 
\cite{donas1}, our selected IMF is a two-component power law with a break
at M=1.8 \msol, with low-mass slope -0.6 and high-mass slope -2.5.
For simplicity we use a model with constant star formation, which has 
constant UV flux after 100 Myr (Landsman, personal
communication), so our integration times are chosen to be 300 Myr.  
This model allows us to approximate the star formation rate as a function 
of the FUV (1520\AA) luminosity as

\[ SFR = \frac{L_{1520}}{4.46\times 10^{39}} M_{\odot} yr^{-1} \] 

This expression gives good agreement with other models 
having similar assumptions ({\em cf} \cite{donas1}), in view of the 
uncertainties present in the approach.  Star formation rates computed 
from this expression for Abell 2246 galaxies, ranging from 1.4 to
 2.9 \msol yr$^{-1}$, are given in  Table~\ref{abonly}.  These values are 
lower limits for actual SFRs, since extinction will decrease the apparent
$L_{1520}$ values.  The expected redshift correction for {\em current} star 
formation rates at z=0.225, as discussed below, is small.

Only a few nearby clusters of galaxies have been observed in any ultraviolet 
bandpass.  \cite{donas2} and \cite{donas3} discuss $\sim$20\arcsec\/ 
resolution images of a local benchmark, the Coma cluster,
in a bandpass $\sim$150\AA\/ wide centered at $\sim$2000\AA, deriving 
a UV luminosity function including 254 galaxies.  It is particularly 
straightforward to compare their observations with our MUV photometry 
of Abell 2246 galaxies, because the central wavelengths of the bandpasses 
in the clusters' rest frames are nearly equal (1955~\AA\/ for \cite{donas2} 
observations of Coma {\em vs} 2033~\AA\/ for our MUV bandpass at Abell 2246) 
and both  observations contained the entire clusters in their respective 
fields of view.

Assuming a distance modulus of 34.8 for Coma based on z = 0.023 
(\cite{thompson}) we use the MUV luminosity function of \cite{donas2} to 
determine that Coma, at the distance of Abell 2246, would have $\sim$12 
UIT-detected galaxies given our limit of m(MUV)$\sim$19.9.  A small richness 
correction derived by scaling the absolute numbers of Abell 2246 galaxies and Coma galaxies 
modifies the number of Coma galaxies seen at the Abell
2246 distance to be $\sim$13-14. This result is consistent with the MUV
data for Abell 2246: while only 5 Abell 2246 galaxies are
{\em uniquely} detected and identified because of the confusion due to 
multiple sources, many more are apparent in the image.  Large numbers
of detectable MUV sources are present in the UIT MUV image of Abell 2246 
because the MUV bandpass, sensitive to stars as late in spectral 
type as F, detects bright elliptical galaxies as well as those with 
current star formation.  

The number of FUV-bright galaxies in Abell 2246 is apparently higher 
than in the Coma Cluster, however.  We may make that comparison using UIT 
Astro-2 FUV observations (1900 sec exposure time; detection limit 
m(FUV)$\sim$19.0) of a field containing the central $\sim$ 40\arcmin\ 
of the Coma cluster.  The two brightest galaxies in the UIT field, 
NGC4858 and IC4040, have UIT-determined m(FUV) of 14.65 and 14.70, 
respectively. Both are spirals and IRAS sources, and are among the six 
brightest Coma galaxies in the MUV according to \cite{donas2}.  
These two galaxies, at the distance of Abell 2246, would have m(FUV) of 
19.65 and 19.70 respectively--slightly fainter than the faintest observed Abell 
2246 galaxy.  Taking into account the redshift correction
to z=0.225 depends critically on stellar content.  SEDs with significant 
young stellar population ({\em e.g.} upper curves in Figure~\ref{colorcolorm}) 
have essentially unchanged FUV flux ($|\Delta m(FUV)|<$0.1), 
while SEDs without the youngest stars become  0.8-1.0 magnitudes fainter at 
z=0.225. In either case, these two galaxies would be fainter than detected 
Abell 2246 galaxies. 
 
We also compare estimates derived from MUV observations of  
the entire Coma cluster by {\cite{donas2}. Coma's MUV luminosity function 
shows only 1 galaxy, NGC 4848 at m(MUV)=13.8, brighter than m(MUV)=14.6. 
(NGC 4848 and other brighter members were outside the UIT Coma field of 
view).  If we assume NGC4848 to have FUV flux and colors indicating   
significant recent star formation, then (FUV--MUV)$\sim-0.5$, and the 
FUV redshift correction is small, resulting in m(FUV)=18.2--18.4, 0.3 mag
brighter than any Abell 2246 galaxy.  If we assume NGC4848 to have the 
colors of normal spiral galaxies, then (FUV--MUV)$\sim$0.0, and NGC4848 
at Coma would have, after redshift correction, m(FUV) = 19.6-19.8, 
approximately equal to the faintest observed Abell 2246 galaxy.  A FUV 
measurement of NGC4848 would resolve this ambiguity; in either case NGC4848
is the only Coma galaxy that is possibly brighter in the FUV, and all others 
are certainly fainter, than Abell 2246 galaxies. It is noteworthy that the 
radial distribution within the cluster of identified Abell 2246 FUV sources is 
similar to Coma's distribution of MUV sources (\cite{donas3}), with 3 of 5 
Abell 2246 galaxies at a radius $\sim$0.33 of the Abell radius (though the 
statistics are inadequate). 

The rarity of FUV-bright Coma cluster galaxies 
is evidence that Abell 2246 has undergone 
significantly more recent star formation than Coma (which nonetheless
shows, in its MUV characteristics, evidence of star formation within the 
past 1 Gyr; {\em cf} \cite{donas3}). This comparison is 
relevant to the Butcher-Oemler effect (\cite{butcher1}; 
\cite{butcher2}) wherein clusters are found to contain fractions of blue 
galaxies (as determined by restframe (B--V)) which increase with redshift.  
The FUV-bright galaxies in Abell 2246 directly show high rates of star 
formation $\sim$1.2 Gyr ago at z=0.225, which are not present to the same
extent at more recent times and smaller redshifts in Coma.  FUV data 
uniquely confirm the nature and scope of this star formation; additional
FUV imaging of clusters such as Abell 2246 with higher sensitivity and 
finer resolution will be be vital in bridging the gap 
between the local neighborhood and the distant universe.



\acknowledgements 

We gratefully acknowledge the contributions made by the many people 
involved in the Astro-2 mission.  We thank Joan Hollis and Joel 
Offenberg for careful and creative image processing and programming help.
  
Funding for the UIT project has been through the Spacelab Office at NASA
under project number 440-51.  RWO gratefully acknowledges NASA support of
portions of this research through grants NAG5-700 and NAGW-2596 to the 
University of Virginia.

\end{document}